\documentstyle[aps,epsfig,float,twocolumn,prl]{revtex}
\newcommand{\be}{\begin{equation}}
\newcommand{\ee}{\end{equation}}

\begin{document}
\twocolumn[\hsize\textwidth\columnwidth\hsize\csname @twocolumnfalse\endcsname
\draft
\title {Decay of Correlations in Fermi Systems at Non-zero Temperature}
\author{M. B. Hastings}
\address{
T-13 and Center for Nonlinear Studies, Los Alamos National
Laboratory, Los Alamos, NM 87545, hastings@lanl.gov 
}
\date{June 7, 2004}
\maketitle
\begin{abstract}
The locality of correlation functions is considered for Fermi systems
at non-zero temperature.  We show that for all short-range, lattice
Hamiltonians, the correlation function of any two fermionic operators
decays exponentially with a correlation length
which is of order the inverse temperature for small temperature.
We discuss applications to numerical simulation of quantum systems
at non-zero temperature.
\vskip2mm
\end{abstract}
]

The locality of correlation functions is a fundamental property of a
quantum system, with great practical importance in
density functional theory simulation of quantum systems\cite{wk}:
if the correlation
functions are short-ranged, then one can develop fast methods $O(N)$ for
simulating a system by solving it in local regions\cite{ordn,ordn2}.  
How, though, can we know {\it a priori} that the correlation functions will 
be short-ranged?
For quantum systems, it was proven recently\cite{lhd} that
the exponential decay in space of connected correlation functions 
follows from the existence
of a gap between the ground and first excited states.
If instead the gap vanishes, as occurs at a quantum phase 
transition\cite{qcrit}, then long-range correlations may occur at zero
temperature, while
slow fluctuations of the order parameters
may give rise to exotic physical properties.

However, another important possibility in which certain correlations will
decay exponentially in space is to
consider a system at non-zero temperature.  In this paper, we prove that
correlation functions of {\it fermionic} operators are short-ranged
at non-zero temperature, and we show that the correlation length is
bounded, for small temperature, by a quantity which scales inversely with the
temperature.
The results are general, and apply to any lattice Hamiltonian 
${\cal H}$ with finite range $R$ as defined below.  There are no
periodicity requirements and no requirements of free or weakly
interacting particles.

Previously, it had been suggested that for non-interacting particles the
correlation length would be inversely proportional to the square-root of
the temperature\cite{bhg}.  However, later it was argued that, for certain
specific systems with free particles in periodic potentials, the
correlation length may be much marger, scaling inversely with
the temperature\cite{iba}.  Thus, the results
in \cite{iba} provide an example showing in this paper we have, in fact,
derived the best possible bound on the scaling
of the correlation length with temperature for small temperature.

It is important that we consider fermionic operators; in contrast, bosonic
operators may develop long-range order at non-zero temperature.  To
define what is meant by fermionic and bosonic operators,
we first define, for each lattice site,
various fermionic and bosonic operators which
act only on that site.  Possible examples of fermionic operators on a site $i$
include electron creation and annihilation
operators, $\psi^{\dagger}_i,\psi_i$; examples of bosonic operators
may include spin operators, phonon operators, etc...
The defining property of an operator on a single site being fermionic or
bosonic is that operators on different sites anticommute with each other if
they are both fermionic, while they commute with each other if at least
one of the operators is bosonic. 
In general, we define an operator $A$ to be fermionic if
$A$ can be written as a sum of products of single-site operators,
where each product includes an {\it odd} number of fermionic single-site
operators.  Thus, $\psi_i$ and $\psi_i \psi^{\dagger}_j \psi_k$ are all allowed
fermionic operators.  

We define the correlation function
at inverse temperature $\beta$ by
$\langle A B \rangle_{\beta}\equiv
Z^{-1} {\rm Tr}[A B \exp(-\beta H)]$, where
$Z={\rm Tr} [\exp(-\beta H)]$.
The trace is taken in the {\it grand} canonical ensemble.
Then, we show that if $A$ and $B$ are fermionic operators separated
by distance $l$ then 
\be
\label{bt}
\langle A B \rangle_{\beta} \leq c ||A|| ||B|| \exp(-l/\xi),
\ee
where $c$ is a constant and, for large $\beta$, the correlation
length $\xi$ is of order
$v \beta$, with $v$ a characteristic velocity of the system.
Here, we define the distance $l$ between operators to
be the shortest distance between any pair of sites $i,j$ where some operator
on site $i$ appears in $A$ and some operator on $j$ appears in $B$.

The finite range condition is defined following 
\cite{lhd,fgv}: we require that we can write ${\cal H}=\sum_{i}
{\cal H}_i$, where $i$ ranges over lattice sites, and where the commutator
$[{\cal H}_i,O]=0$ for any operator $O$ which only acts on sites $j$ which
are more than a distance $R$ from site $i$, and where, for some constant
$J$, $||{\cal H}_i||\leq J$ for all $i$.  These conditions are sufficient to
enforce the requirement of a finite group velocity below.  As an example
of the finite range, a hopping term
$\psi^{\dagger}_i \psi_j$ in the Hamiltonian ${\cal H}_i$ is allowed if
site $j$ is within range $R$ of site $i$.

To prove Eq.~(\ref{bt}), we first
express the correlation function in terms of anti-commutators using
an integral representation.  Then,
we bound the anti-commutators using the finite group velocity\cite{loc}.

{\it Integral Representation of Correlation Function---}
Let $A$ have matrix elements $A_{ij}$ in a basis of eigenvectors of
${\cal H}$, where states $i,j$ have energies $E_i,E_j$ respectively.
Let $(A_{\omega})_{ij}=A_{ij}\delta(E_i-E_j-\omega)$.  Note that
$A=\int {\rm d}\omega\, A_{\omega}$.

The expectation value $\langle A_{\omega} B \rangle_{\beta}
=Z^{-1} \sum_{i,j} \delta(E_i-E_j-\omega)
A_{ij}B_{ji} \exp(-\beta E_i)$.  Similarly,
$\langle B A_{\omega} \rangle_{\beta}=
Z^{-1} \sum_{i,j} \delta(E_i-E_j-\omega)
B_{ji}A_{ij} \exp(-\beta E_j)$.  
However, $\delta(E_i-E_j-\omega)
\exp(-\beta E_j)=\delta(E_i-E_j-\omega)\exp(-\beta E_i)\\ \exp(\beta \omega)$.
Thus,
$\langle B A_{\omega} \rangle_{\beta}=\langle A_{\omega} B \rangle_{\beta}
\exp(\beta \omega)$.  
Therefore,
\begin{eqnarray}
\label{ac}
\langle A_{\omega} B \rangle_{\beta}=\frac{1}{1+\exp(\beta \omega)}
\langle \{A_{\omega},B\} \rangle_{\beta}.
\end{eqnarray}

It was essential to use the grand canonical ensemble to derive this
relation: if instead
we worked in the canonical ensemble with a fixed number, $N$, of particles, then
in the definition of 
$\langle A_{\omega} B \rangle_{\beta}$ as a sum over states $i,j$ we would
require that state $i$ had $N$ particles while in the definition of
$\langle B A_{\omega} \rangle_{\beta}$ we would instead require that state
$j$ had $N$ particles.  If the operator $A$ changes the number of
particles we would not be able to relate these two expectation values as
simply as was done above.

Next, we use $(1+\exp(\beta \omega))^{-1}=1/2-\beta^{-1} \sum_{n {\rm odd}}
(\omega-i n \pi/\beta)^{-1}$, where the sum ranges over all positive and
negative {\it odd} $n$.  For $n>0$, we have
$(\omega-i n \pi/\beta)^{-1}=i\int_{0}^{\infty} {\rm d}t
\exp[-(i\omega+n\pi/\beta)t]$.  Similarly, for $n<0$, we have
$(\omega-i n \pi/\beta)^{-1}=-i\int_{0}^{\infty} {\rm d}t
\exp[(i\omega+n\pi/\beta)t]$.  Thus,
\begin{eqnarray}
\label{sum}
\frac{1}{1+\exp(\beta \omega)}=1/2+\\ \nonumber \frac{i}{\beta}
\sum_{n {\rm odd},n>0}
\int_{0}^{\infty} {\rm d}t\,
\exp(-n \pi t/\beta) 
 (\exp(i\omega t)-\exp(-i\omega t)),
\end{eqnarray}
where now the sum ranges only over positive, odd $n$.

Now, we define the time evolution of operators by $A(t)=\exp(i{\cal H} t)
A \exp(-i{\cal H} t)$.  Thus, $A_{\omega}(t)=\exp(i \omega t) A_{\omega}$.
Combining Eqs.~(\ref{ac},\ref{sum}), we get
\begin{eqnarray}
\label{tint}
\langle A_{\omega} B \rangle_{\beta}=1/2 \langle \{A_{\omega},B\} 
\rangle_{\beta}+
\\ \nonumber
\frac{i}{\beta}
\int_{0}^{\infty} {\rm d}t\,
\frac{\exp(-\pi t/\beta)}{1-\exp(-2\pi t/\beta)}
\langle \{A_{\omega}(t)-A_{\omega}(-t),B\} \rangle_{\beta},
\end{eqnarray}
where we have used 
$\sum_{n {\rm odd},n>0} \exp(-n \pi t/\beta)=
\exp(-\pi t/\beta)/(1-\exp(-2\pi t/\beta))$.
Finally, we integrate Eq.~(\ref{tint}) over $\omega$ to get
\begin{eqnarray}
\label{tint2}
\langle A B \rangle_{\beta}=1/2 \langle \{A,B\} 
\rangle_{\beta}+
\\ \nonumber
\frac{i}{\beta}
\int_{0}^{\infty} {\rm d}t\,
\frac{\exp(-\pi t/\beta)}{1-\exp(-2\pi t/\beta)}
\langle \{A(t)-A(-t),B\} \rangle_{\beta}.
\end{eqnarray}

{\it Finite Group Velocity---}
In \cite{fgv,lhd}, it was proven that if we have a Hamiltonian obeying the
finite range condition above, then for any two {\it bosonic} operators $A,B$,
separated in space by a distance $l$, one can bound the operator norm of
the commutator: $||[A(t),B]|| \leq ||A|| ||B|| \sum_j g(t,l_j)$, where the
sum ranges over sites $j$ which appear in operator $B$ and $l_j$ 
is the distance
from $j$ to the closest site $i$ in the operator $A$.  Here $g(t,l_j)$ is
a function with the following properties: (1) $g$ is symmetric in $t$ so
that $g(t,l)=g(-t,l)$; (2) $g(t,l)\leq (t/t') g(t',l)$ for $t<t'$ and $t,t'>0$;
and (3) there exists a constant $c_1$ such that $g(c_1 l,l)$ is exponentially
decaying in $l$ for large $l$ with some correlation length $\xi_C$.  Defining
$v=c_1^{-1}$, we can view $v$ as a characteristic velocity of the system.

While in \cite{fgv,lhd} only bosonic operators were considered, for
{\it fermionic} operators, a similar bound can be proven:
$||\{A(t),B\}||\leq ||A|| ||B|| 
\sum_j g(t,l_j)$.  This proof can be proven following the same steps as
in \cite{lhd} with commutators replaced by anti-commutators throughout.

We now combine the bound on anti-commutators with Eq.~(\ref{tint2}).  The
physics is as follows: Eq.~(\ref{tint}) gives the correlation function
as an integral over time.  However, for times $t$ large compared to 
$\beta$, the integral is cut off by the exponential; however, for times
$t$ of order $\beta$ or less, the
anti-commutator of Eq.~(\ref{tint}) is small for $l$ large compared to
$v\beta$.

More precisely, note that since $A,B$ are separated by a distance $l$,
then $\{A,B\}=0$.
Thus, 
\begin{eqnarray}
\label{twi}
|\langle A B \rangle_{\beta}|\leq \\ \nonumber
|\frac{i}{\beta}
\int_{0}^{c_1 l} {\rm d}t\,
\frac{\exp(-\pi t/\beta)}{1-\exp(-2\pi t/\beta)}
\langle \{A(t)-A(-t),B\} \rangle_{\beta}|+ \\ \nonumber
|\frac{i}{\beta}
\int_{c_1 l}^{\infty} {\rm d}t\,
\frac{\exp(-\pi t/\beta)}{1-\exp(-2\pi t/\beta)}
\langle \{A(t)-A(-t),B\} \rangle_{\beta}|.
\end{eqnarray}.

Note that $|\langle \{A(t)-A(-t),B\} \rangle_{\beta}|\leq 4||A|| ||B||$.
Thus, the integral over times $t>c_1 l$ in Eq.~(\ref{twi}) is bounded in
absolute value by
$4 ||A|| ||B|| \beta^{-1}\int_{c_1 l}^{\infty} \exp(-\pi t/\beta)/(1-
\exp(-2\pi t/\beta))\leq 
4 ||A|| ||B|| \pi^{-1} \exp(-\pi c_1 l/\beta)/
(1-\exp(-2\pi c_1 l/\beta))$.

For times $t<c_1 l$, we use the bound on the anti-commutator:
$|\langle \{A(t),B\} \rangle_{\beta}|\leq 
||\{A(t),B\}||\leq [t/(c_1 l)] \sum_j g(c_1 l,l_j)$.  Also, 
$\exp(-\pi t/\beta)/(1-\exp(-2 \pi t/\beta))\leq \beta/(2 \pi t)$.
Thus, the integral over times $t<c_1 l$ in Eq.~(\ref{twi}) is bounded in
absolute value by $||A|| ||B|| 
\sum_j g(c_1 l,l_j)\int_0^{c_1 l} {\rm d}t (2 \pi c_1 l)^{-1}=
(2\pi)^{-1} ||A|| ||B|| \sum_j g(c_1 l,l_j)$.

Therefore,
\begin{eqnarray}
\label{bf}
|\langle A B \rangle_{\omega}| \leq
||A|| ||B|| \Bigl( 
\frac{\sum_j g(c_1 l,l_j)}{2\pi}+
\frac{4}{\pi} \frac{\exp(-\pi c_1 l/\beta)}{1-\exp(-2\pi c_1 l/\beta)}\Bigr).
\end{eqnarray}
Since $l_j\geq l$, $g(c_1 l,l_j)$ decays exponentially for large $l$
as $\exp(-l/\xi_C)$.  Also,
$\exp(-\pi c_1 l/\beta)$ decays exponentially with correlation
length $v \beta/\pi$.  Thus, the correlation length $\xi$ is bounded by
the
maximum of $\xi_C$ and $v\beta/\pi$, so that for small temperature
$\xi\leq v\beta/\pi$.
Thus, Eq.~(\ref{bt}) follows.

It is interesting to compare Eq.~(\ref{bf}) to the result found in \cite{lhd}
for systems in the ground state
with a gap $\Delta E$ between the ground and first excited
states.  There, the resulting correlation length was the maximum of
$\xi_C$ and $2 v/\Delta E$.

{\it Importance of Lattice Structure---}
This bound was derived for systems with a lattice structure.  To illustrate
the importance of a lattice, consider free fermions on a lattice
in one dimension, with
Hamiltonian $t \psi^{\dagger}_i \psi_{i+1}+h.c.$  The Fermi velocity depends
on filling, and is maximum at half-filling.  Thus, the Fermi velocity
can be bounded, and is at most of order $t$.  Now, consider free fermions in
free space one dimension, with Hamiltonian $\psi^{\dagger}(x) 
[\partial^2/(2m)] \psi(x)$.  The Fermi velocity again depends on the particle
density, but can be increased without limit as the particle density is 
increased.

This is the physical reason why,
for systems which are not on a lattice, one cannot (without
knowing more about the density and other details of the system) bound the
anti-commutator.  One cannot, from the Hamiltonian alone, provide a velocity
$v$ such that the anti-commutator $\{A(t),B\}$ is small for $t<l/v$.

{\it Discussion---}
We have proven a bound on the correlation of fermionic operators
on non-zero temperature, Eq.~(\ref{bf}).  There are specific examples\cite{iba}
which show that this bound on the scaling of
the correlation length with temperature is the best possible (although
the prefactor $v/\pi$ in the correlation length $\xi=v\beta/\pi$ might not
be the best possible).
Similarly, while it had been shown that for periodic, non-interacting
insulators in one-dimension the
correlation length scales inversely with the square-root of the gap\cite{wke},
there are again specific examples of non-interacting systems\cite{iba} which
show that the bound\cite{lhd} on the scaling of the correlation length with
the gap is also the best possible.  The case of the canonical, as opposed
to grand canonical, ensemble is an interesting future problem.

In addition to the basic interest in this result in quantum statistical
mechanics, this result is of importance in quantum simulation using
density functional theory.  Also, the integral representation, Eq.~(\ref{tint2})
may prove useful as a means of computing the density matrix for
free fermions at non-zero temperature; for these systems, if $A(0)$ is
the fermionic creation operator on a single site and $B$ is a fermionic
annihilation operator, then $A(t)$ is a sum
of fermionic creation operators and the anti-commutator is trivial to
compute.  This is a matter for future research.

{\it Acknowledgements---}
I thank A. Niklasson for useful discussions.
This work was supported by DOE contract W-7405-ENG-36.


\begin{thebibliography}{99}
\vspace{-15mm}

\bibitem{wk} W. Kohn, Phys. Rev. Lett. {\bf 76}, 3168 (1996).

\bibitem{ordn} W. Yang, Phys. Rev. Lett. {\bf 66}, 1438 (1991).

\bibitem{ordn2} S. Goedecker and M. Teter, Phys. Rev. B {\bf 51}, 9455 (1995).

\bibitem{lhd} M. B. Hastings, Phys. Rev. B {\bf 69} 104431 (2004).

\bibitem{qcrit} S. Sachdev, {\it Quantum Phase Transitions},
(Cambridge University Press, New York, 1999).

\bibitem{bhg} R. Baer and M. Head-Gordon, Phys. Rev. Lett. {\bf 79}, 3962 
(1997).

\bibitem{iba} S. Ismail-Beigi and T. A. Arias, Phys. Rev. Lett. {\bf 82},
2127 (1999).

\bibitem{loc} M. B. Hastings, preprint cond-mat/0405587.

\bibitem{fgv} E. Lieb and D. Robinson, Commun. Math. Phys. {\bf 28},
251 (1972).

\bibitem{wke} W. Kohn, Phys. Rev. {\bf 115}, 809 (1959).
\end{thebibliography}
\end{document}